\documentclass[sigconf,natbib=true,anonymous=false]{acmart}

\AtBeginDocument{%
  \providecommand\BibTeX{{%
    \normalfont B\kern-0.5em{\scshape i\kern-0.25em b}\kern-0.8em\TeX}}}

\copyrightyear{2023}
\acmYear{2023}
\setcopyright{acmlicensed}\acmConference[SIGIR '23]{Proceedings of the 46th International ACM SIGIR Conference on Research and Development in Information Retrieval}{July 23--27, 2023}{Taipei, Taiwan}
\acmBooktitle{Proceedings of the 46th International ACM SIGIR Conference on Research and Development in Information Retrieval (SIGIR '23), July 23--27, 2023, Taipei, Taiwan}
\acmPrice{15.00}
\acmDOI{10.1145/3539618.3591664}
\acmISBN{978-1-4503-9408-6/23/07}

\usepackage{footmisc}
\usepackage{subcaption,siunitx,booktabs}
\usepackage{caption}
\usepackage{bbm}
\usepackage{multirow}
\usepackage{balance}
\usepackage{enumitem}
\usepackage{wrapfig}
\newcommand{\zsy}[1]{\iffalse #1 \fi{\color{blue} \textbf{[ZSY]}}}

\begin{document}
\title{DisCover: Disentangled Music Representation Learning for Cover Song Identification}

\author{Jiahao Xun}
\affiliation{%
  \institution{Zhejiang University}
  \city{Hangzhou}
  \country{China}}
\email{jhxun@zju.edu.cn}

\author{Shengyu Zhang$^{\dagger}$}
\affiliation{%
  \institution{Zhejiang University}
  \city{Hangzhou}
  \country{China}}
\email{sy_zhang@zju.edu.cn}

\author{Yanting Yang}
\affiliation{%
  \institution{Zhejiang University}
  \city{Hangzhou}
  \country{China}}
\email{yantingyang@zju.edu.cn}

\author{Jieming Zhu}
\affiliation{%
  \institution{Huawei Noah's Ark Lab}
  \city{Shenzhen}
  \country{China}}
\email{jiemingzhu@ieee.org}

\author{Liqun Deng}
\affiliation{%
  \institution{Huawei Noah's Ark Lab}
  \city{Shenzhen}
  \country{China}}
\email{dengliqun.deng@huawei.com}

\author{Zhou Zhao$^{\dagger}$}
\affiliation{%
  \institution{Zhejiang University}
  \city{Hangzhou}
  \country{China}}
\email{zhaozhou@zju.edu.cn}

\author{Zhenhua Dong}
\affiliation{%
  \institution{Huawei Noah's Ark Lab}
  \city{Shenzhen}
  \country{China}}
\email{dongzhenhua@huawei.com}

\author{Ruiqi Li}
\affiliation{%
  \institution{Zhejiang University}
  \city{Hangzhou}
  \country{China}}
\email{rickyli@zju.edu.cn}

\author{Lichao Zhang}
\affiliation{%
  \institution{Zhejiang University}
  \city{Hangzhou}
  \country{China}}
\email{zju_zlc@zju.edu.cn}

\author{Fei Wu}
\affiliation{%
  \institution{Zhejiang University}
  \city{Hangzhou}
  \country{China}}
\email{wufei@zju.edu.cn}

\renewcommand{\shortauthors}{Jiahao Xun and Shengyu Zhang, et al.}
\newcommand{\etal}{\textit{et al}.}
\newcommand{\ie}{\textit{i}.\textit{e}.}
\newcommand{\eg}{\textit{e}.\textit{g}.}
\newcommand{\vpara}[1]{\vspace{0.05in}\noindent\textbf{#1 }}

\begin{abstract}
In the field of music information retrieval (MIR), cover song identification (CSI) is a challenging task that aims to identify cover versions of a query song from a massive collection. Existing works still suffer from high intra-song variances and inter-song correlations, due to the entangled nature of version-specific and version-invariant factors in their modeling. In this work, we set the goal of disentangling version-specific and version-invariant factors, which could make it easier for the model to learn invariant music representations for unseen query songs. We analyze the CSI task in a disentanglement view with the causal graph technique, and identify the intra-version and inter-version effects biasing the invariant learning. To block these effects, we propose the disentangled music representation learning framework (DisCover) for CSI. DisCover consists of two critical components: (1) Knowledge-guided Disentanglement Module (KDM) and (2) Gradient-based Adversarial Disentanglement Module (GADM), which block intra-version and inter-version biased effects, respectively. KDM minimizes the mutual information between the learned representations and version-variant factors that are identified with prior domain knowledge. GADM identifies version-variant factors by simulating the representation transitions between intra-song versions, and exploits adversarial distillation for effect blocking. Extensive comparisons with best-performing methods and in-depth analysis demonstrate the effectiveness of DisCover and the and necessity of disentanglement for CSI.
\renewcommand{\thefootnote}{\fnsymbol{footnote}}
\footnotetext[2]{Corresponding Authors.}
\renewcommand{\thefootnote}{}
\footnote{The source code will be available at \url{https://gitee.com/mindspore/models}.}
\end{abstract}




\maketitle

\section{Introduction}
\begin{figure}[htp] \centering
    \includegraphics[width=\columnwidth]{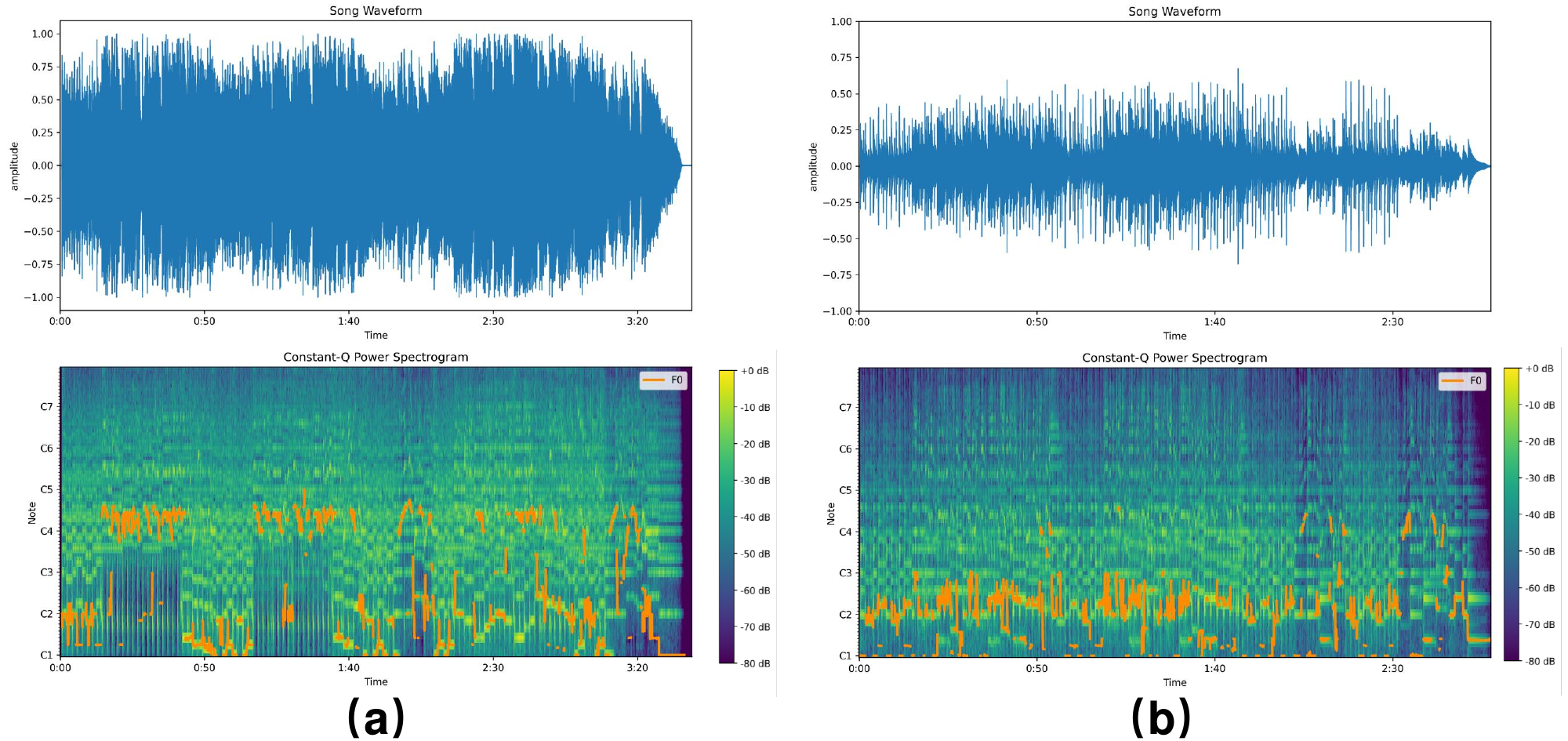}
    \caption{ An illustration of the data distribution in CSI task with the raw waveform (top) and CQT spectrogram (bottom). (a) and (b) are two different cover version for the same song "Don't Let It Bring You Down" in Covers80 dataset. 
	}
\label{fig:data}
 \end{figure}

Nowadays, online digital music platforms, such as  Spotify
and Apple Music
contain a massive number of music tracks for consumption, intensifying the need of music retrieval techniques for discovering related songs. One of the key techniques for music discovery is cover song identification (CSI), which aims to retrieve the cover versions from a music collection given a query song. Specially, a cover version/song is an alternative interpretation of the original version with different musical facets (\eg\ timbre, key, tempo, or structure). In real-world scenarios, the music collection can be massive and rapidly updated, potentially amplifying the intra-song variances (\textit{c.f.} Figure \ref{fig:data}) and inter-song correlations.
These characteristics drive the CSI problem hard to handle due to the ubiquitous spurious correlations among songs of different collections.
Intuitively, CSI requires a fine-grained analysis of music facets and semantics such that the intra-song correlations and inter-song differences can be adequately distinguished.

%

Recently, with the development of artificial intelligence in other domains \cite{wang2021clicks, wang2021deconfounded, wang2021denoising, wang2022causal, liu2022pseudo, zhang2022hifidenoise, yin2021simulslt, yin2022mlslt,GASLT}, deep learning based CSI models have presented superior performance compared with traditional sequence matching methods \cite{marolt2006mid, ellis2007identifyingcover, serra2008chroma, serra2009cross}. Most of them \cite{xu2018key, yu2019temporal, yu2020learning, du2021bytecover,du2022bytecover2, DBLP:conf/interspeech/HuZLJWKZJ22}  treat CSI as a classification task and utilize CNN-based architecture for music content understanding.
Furthermore, some state-of-the-art works \cite{du2021bytecover,du2022bytecover2,DBLP:conf/interspeech/HuZLJWKZJ22} explore metric learning techniques to narrow the gap between different cover versions of the same song and simultaneously expand the distance among the different version groups. 
Despite the significant advances made by these methods, we argue that song-specific and song-sharing musical factors are highly entangled in their modeling, thus being inadequate to distinguish unseen cover versions and songs. For example, as shown in Figure \ref{fig:data}, the testing cover version (b) of the query song (a) shows significant differences in pitch/F0 (the orange curve in the spectrogram), timbre, or rhythm. If the model is unable to disentangle these factors and identifies them as version-variant, it might fail to generalize on this testing version and identify it as negative ones. On the other hand, if the model fails to disentangle and recognize version-invariant factors, it might falsely correlate some other songs with the given query based on the high similarity of version-specific factors. To bridge the gap, we set the goal of explicitly disentangling version-variant and version-invariant factors and thus learning invariant musical representations for unseen cover song identification.

\begin{figure}[] 
\subcaptionbox{Model’s perspective\label{subfig:causal_graph_a}}[0.45\columnwidth]{
\includegraphics[width=0.45\columnwidth]{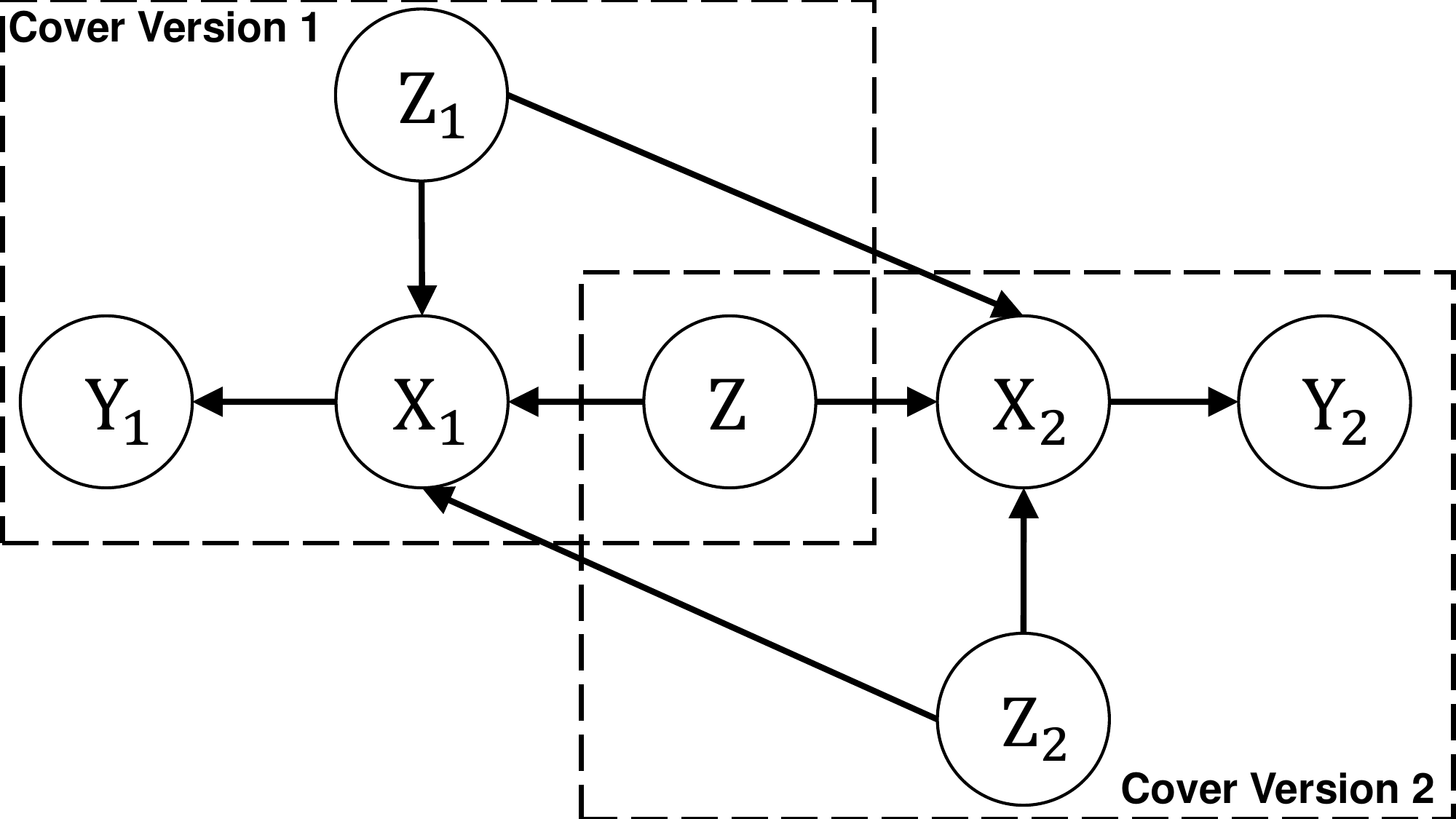}
}\quad
\subcaptionbox{Searcher's perspective\label{subfig:causal_graph_b}}[0.45\columnwidth]{
\includegraphics[width=0.45\columnwidth]{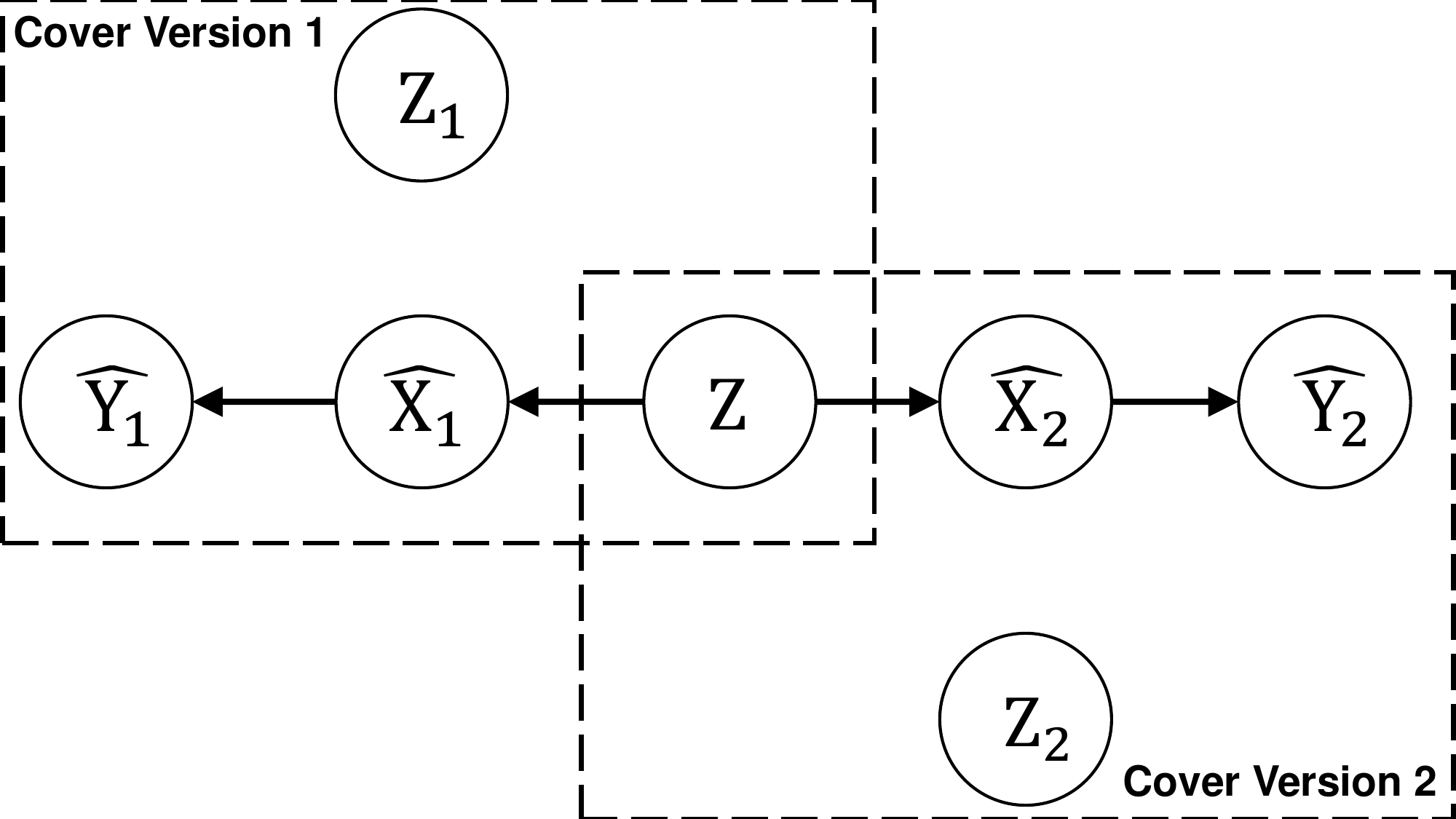}
}

\caption{Causal graph of CSI from different perspectives.}
\label{fig:causal_graph}
\vspace{-0.5cm}
\end{figure} 

To better understand the underlying mechanism of the disentanglement in CSI task, we resort the causal graph technique \cite{pearl2009causality} for illustration (\textit{c.f.} Figure \ref{fig:causal_graph}).  
The nodes denote cause or effect factors. An edge $A \rightarrow B$ means the $A$ can directly affect $B$.
\begin{itemize}[leftmargin=*]
	\item Firstly, we illustrate the causal graph from the model’s perspective in
Figure \ref{fig:causal_graph}(\subref{subfig:causal_graph_a}): $Z_i$ denotes the set of factors that are specific to the $i$-th cover version and are mostly version-variant. $X_i$ denotes the learned musical representation of the $i$-th cover version. $Z$ denotes the set of version-invariant factors.
$Y_i$ denotes the retrieval results (\eg\ a candidate playlist) given the learned representation $X_i$. Intuitively, during the co-training of various cover versions where different factors are highly entangled, $Z_1$ could have a direct effect on $X_1$ and also $X_2$, which is the musical representation of the second cover version.
Therefore, $Y_1$ and $Y_2$ will be indirectly affected through causal path $Z_1 \rightarrow X_1 \rightarrow Y_1$ and $Z_1 \rightarrow X_2 \rightarrow Y_2$ respectively, which will lead to spurious correlations and mismatching during unseen cover song identification. 
    \item Secondly, as illustrated in Figure \ref{fig:causal_graph}(\subref{subfig:causal_graph_b}), we further consider the causal graph from searcher's perspective. It is a relatively ideal causal graph that $\hat{X_i}$ is only affected by $Z$. In other words, version information has no effect on the learned music representation, such that intra-song versions can be adequately distinguished from the others. 

\end{itemize}

In this work, we aim to develop a disentanglement framework that could realize the transition of models' underlying behavior from Figure \ref{fig:causal_graph}(\subref{subfig:causal_graph_a}) to Figure \ref{fig:causal_graph}(\subref{subfig:causal_graph_b}) for debiased and effective cover song identification. We identify two critical challenges in achieving disentanglement in CSI: (1) Mitigating the negative effect from cover information and extracting the commonness for the versions (cutoff $Z_i \rightarrow X_i \rightarrow Y_i$), which aims to make the model more focused on the version-invariant factors $Z$ and learn invariant representations for different cover versions. (2) Identifying the differences between versions and alleviating the negative transfer (cutoff $Z_i \rightarrow X_j \rightarrow Y_j$), which attempts to bridge the intra-group gap and avoid biased representation learning. 
It is non-trivial to block paths $Z_i \rightarrow X_i \rightarrow Y_i$ and $Z_j \rightarrow X_i \rightarrow Y_i$ due to the implicit nature of version-specific factors $Z_i, Z_j$ and the effects in deep neural networks. In this regard, we introduce a disentanglement module for identifying version-specific factors, followed by an effect-blocking module for learning invariant representations. As for the path $Z_i \rightarrow X_i \rightarrow Y_i$, disentangling $Z_i$ is challenging without supervision signals since different factors (\eg\ F0 and timbre) in raw music are highly entangled. In this regard, we introduce prior domain knowledge as guidance for disentanglement. As for the path $Z_j \rightarrow X_i \rightarrow Y_i$, the challenge lies in how to identify the factors $Z_j$ in the $j$-th sample that could affect the representation learning of $X_i$. Intuitively, we regard the modified factors during the transition from $X_j$ to $X_i$ as version-variant factors that are critical to $X_i$ in the $j$-th sample.

Technically, we propose a \textbf{\textit{Dis}}entangled music representation learning framework for \textbf{\textit{Cover}} song identification, denoted as DisCover, which encapsulates two key components: (1) Knowledge-guided Disentanglement Module (\textbf{KDM}) and (2) Gradient-based Adversarial Disentanglement Module (\textbf{GADM}) for blocking biased effects $Z_i \rightarrow X_i \rightarrow Y_i$ and $Z_j \rightarrow X_i \rightarrow Y_i$, respectively. KDM employs off-the-shelf music feature extractors as the domain knowledge for disentanglement, and minimizes the mutual information (MI) between the learned representations and version-variant factors. GADM identifies version-variant factors by simulating the representation transitions between intra-song versions and adopting gradient-based adaptive masking. Since the discrete-valued mask might distort the continuity of representations in the hypersphere, it would be less effective to use MI to measure the effect $Z_j \rightarrow X_i \rightarrow Y_i$. Instead, GADM incorporates an adversarial distillation sub-module for distribution-based effect blocking.

The main highlights of this work are summarized as follows:
\begin{itemize}[leftmargin=*]
	\item We analyze the cover song identification problem in a disentanglement view with causal graph, a powerful tool but is seldom used in the community. We identify the bad impact of version-variant factors with two effect paths that needed to be blocked.
	\item We propose the DisCover framework that disentangles version-variant factors among intra-song versions and blocks two biased effect paths via knowledge-guided MI minimization and gradient-based adversarial distillation.
	\item We conduct in-depth experimental analyses along on both quantitative and qualitative results, which have demonstrated the effectiveness and necessity of disentanglement for CSI.

\end{itemize}

\section{RELATED WORK}
\subsection{Cover Song identification}
With the increasing amount of music data on the Internet, cover song identification (CSI) has long been a popular task in the music information retrieval community. CSI aims to retrieve the cover versions of a given song in a dataset, which can also be seen as measuring the similarity between music signals without meta-information (\eg, title, author, genre). 
Specifically, meta-information might ease the problem but also introduce spurious correlations that many different songs have quite similar or even the same short title. Moreover, users humming the query songs might not necessarily know/provide the meta-information. 
Overall, CSI as a challenging task  has long attracted lots of researchers due to its potential applications in music representation learning \cite{yao2022contrastive,jiang2020transformer}, retrieval \cite{yu2019deep,muller2018cross,simonetta2019multimodal} and recommendation \cite{hansen2020contextual,pereira2019online}. However, those cover songs may differ from the original song in key transposition, speed change, and structural variations, which challenges identifying the cover song. To solve these problems, \cite{serra2009cross} developed music sequences alignment algorithms for version identification by measuring the similarity between time series, 
 and \cite{ellis2012large} generated fixed-length vectors for cover song identification. In addition, deep learning approaches are introduced to CSI. For instance, CNNs are utilized to measure the similarity matrix \cite{chang2017audio} or learn features \cite{qi2017audio,xu2018key,yu2019temporal,doras2019cover}. On this basis, TPPNet \cite{yu2019temporal} uses a temporal pyramid pool to extract information at a different scale. CQTNet \cite{yu2020learning} proposes a special CNN architecture to extract musical representations and train the network through classification strategies. Although these methods have made significant progress, they ignore the entanglement of cover song representations and may incorrectly correlate some other songs with a given query. Thus, we propose a framework that disentangles version-variant factors among intra-song versions.
\subsection{Disentangled Representation Learning}
Disentangled representation learning (DRL) focuses on encoding data points into separate independent embedding subspaces, where different subspaces represent different data attributes. To prevent information leakage from each other, the correlation between the two embedding parts is still required to be reduced. Some correlation-reducing methods mainly focus on Mutual Information (MI) minimization, where MI is a fundamental measure of the dependence between two random variables. To accurately estimate MI upper bound, CLUB \cite{cheng2020club} bridges mutual information estimation with contrastive learning. This method has gained a lot of attention and applications in scenarios such as domain adaption, style transfer, and causal inference. For instance, IDE-VC \cite{yuan2021improving} and VQMIVC \cite{wang2021vqmivc} achieves proper disentanglement of speech representations. MIM-DRCFR \cite{cheng2022learning}  learns disentangled representations for counterfactual regression. In addition, as analyzed in \cite{zhou2016learning,selvaraju2017grad,chattopadhay2018grad}, the gradients of the final predicted score convey the task-discriminative information, which correctly identifies the task-relevant features. For instance, Grad-CAM \cite{selvaraju2017grad} visualizes the importance of each class by leveraging the gradient information. On the basis of this, ToAlign \cite{wei2021toalign}  decomposes a source feature into a task-relevant one and a task-irrelevant one for performing the classification-oriented alignment. RSC \cite{huang2020self}  discards the task-relevant representations associated with the higher gradients. DropClass \cite{chu2021learning}  uses gradient information to extract class-specific information from the entangled feature map. However, most of these works learn to disentangle representations from a single perspective. This paper blocks two biased effect paths via knowledge-guided MI minimization and gradient-based adversarial distillation.

 \subsection{Music Representation Learning}
An effective musical representation is essential for learning different music-related tasks, such as music classification \cite{choi2017convolutional,lee2017sample,pons2019musicnn,van2014transfer,xu2018key,wu2021multi}, cover song identification \cite{yu2019temporal,yu2020learning,xu2018key,yesiler2020accurate}, music generation \cite{huang2022singgan,huang2021multi,liu2022diffsinger,ren2020popmag}. Most of them rely on large amounts of labeled datasets to learn music representations. As the labeled datasets on which supervised learning methods require extensive manual labeling, it is often costly and time-consuming, leading to limitations in the performance of supervised learning methods. For this reason, some audio researchers have adopted a self-supervised learning approach to learning musical representations \cite{zhu2021musicbert,yao2022contrastive,spijkervet2021contrastive,saeed2021contrastive}. For example, MusicBERT \cite{zhu2021musicbert} models music self-representation with a multi-task learning framework. PEMR \cite{yao2022contrastive} proposes a positive-negative frame mask for music representation with contrastive learning. Many approaches to music representation learning focus on key pieces of music, while CSI focuses more on the whole song.

\begin{figure}[] 
\subcaptionbox{Cutoff $\mathbf{Z_i \rightarrow X_i}$\label{subfig:causal_intervention_a}}[0.45\columnwidth]{
\includegraphics[width=0.45\columnwidth]{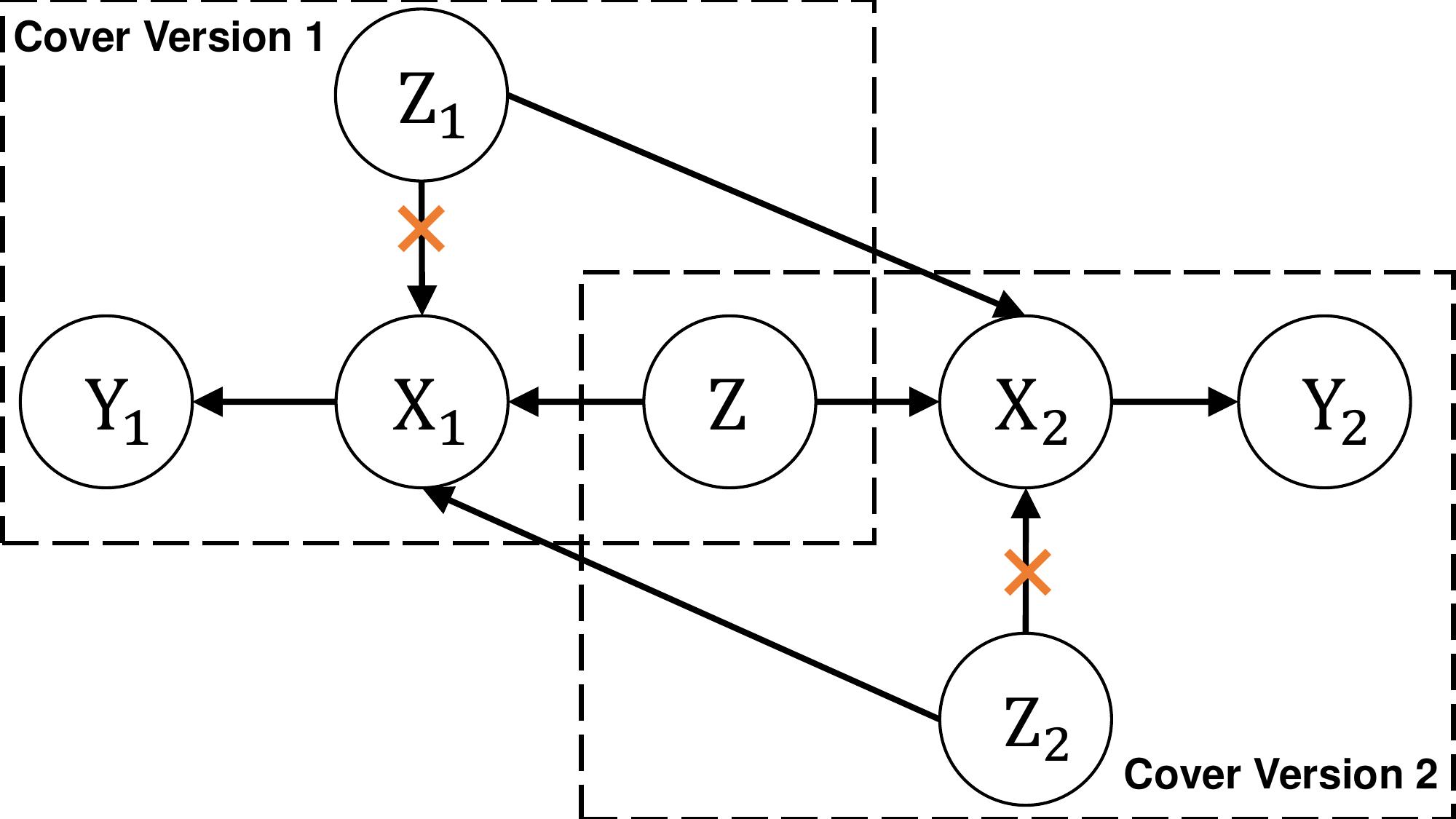}
}\quad
\subcaptionbox{Cutoff $\mathbf{Z_i \rightarrow X_j}$\label{subfig:causal_intervention_b}}[0.45\columnwidth]{
\includegraphics[width=0.45\columnwidth]{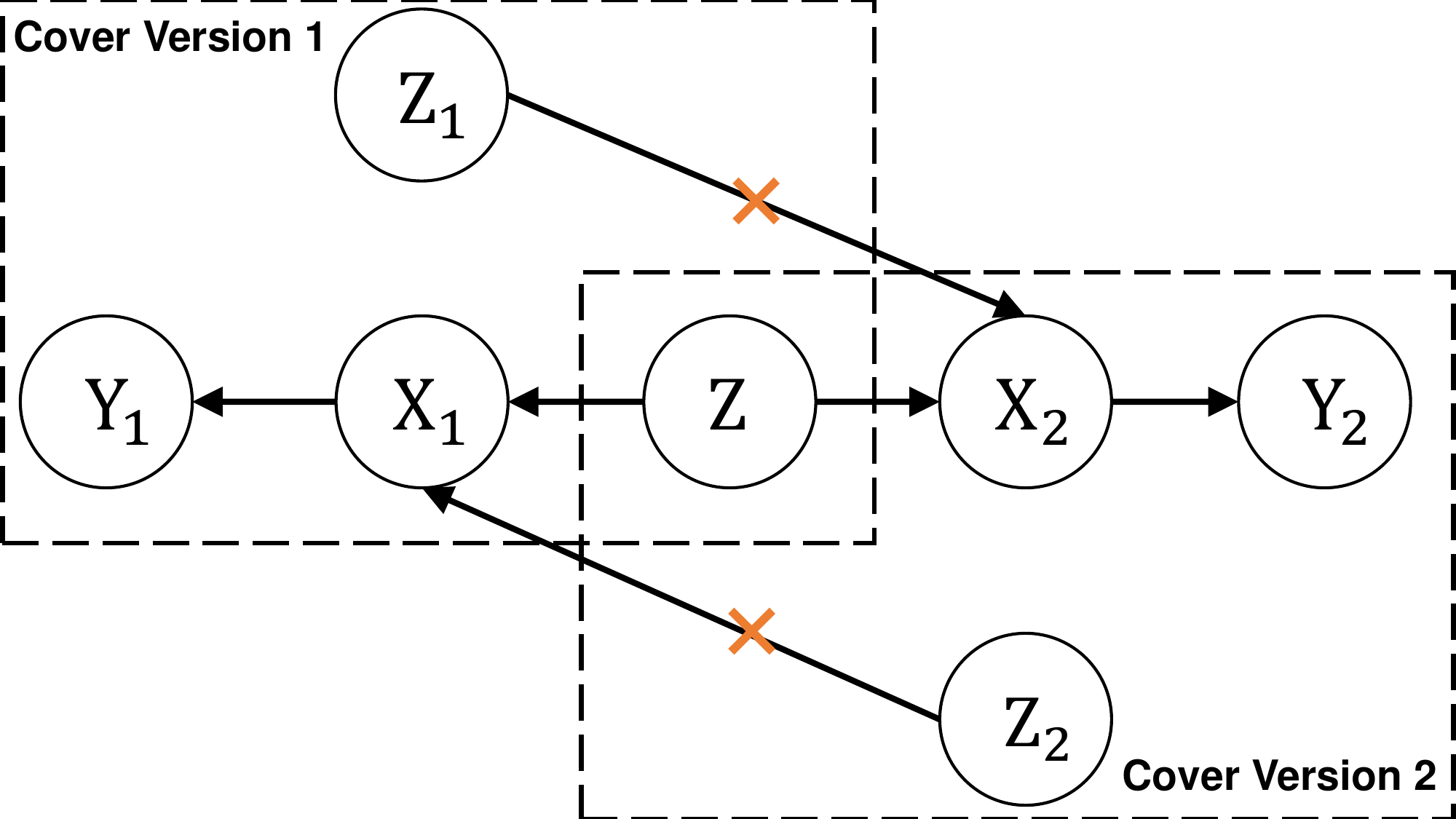}
}

\caption{Interventions on causal graph of DisCover from the perspective of modelling.}
\label{fig:causal_intervention}
\vspace{-0.5cm}
\end{figure} 

\section{PROPOSED METHOD}
\subsection{PRELIMINARIES}
\vpara{\textbf{Problem Formulation.}}
Following the common practice in modern cover song identification task \cite{yu2019temporal,xu2018key}, we formulate cover song identification as an information retrieval problem and specifically focus on music representation learning. We use $q$ to denote one query song and $\mathcal{S} = \{ s_i \}_{i=1,\dots,|\mathcal{S}|}$ to denote the song collections on an online music platform.

Given the query $q$, cover song identification aims to retrieve the most similar candidates $\mathcal{C} = \{ c_i \}_{i=1,\dots,k}$ from the song collections $\mathcal{S}$ in a top-k manner. A deep learning-based CSI model $f(\cdot)$ encodes the $q$ and $s_i$ into the fixed dimension representation $\boldsymbol{q}$ and $\boldsymbol{s}_i$ separately. Then we use cosine distance to calculate the similarity for all the pairs $P = \{(\boldsymbol{q}, \boldsymbol{s}_i)\}_{i=1,\dots,|S|}$. During testing and serving, top-k candidates $\mathcal{C}$ will be ranked by the similarity and displayed on the music platform in a position consistent with the rank.

\vpara{\textbf{Prior Knowledge Selection.}} \label{sec:knowledge}
There are usually multiple variations of musical facets for the cover version, such as timbre, key, tempo, timing, or structure \cite{JoanSerr2010AudioCS}. Hence it meets a problem of how to select the appropriate musical facets as expert knowledge. Inspired by the common practice in the disentanglement-based voice conversion \cite{qian2020unsupervised, wang2021vqmivc} and singing voice synthesis \cite{choi2022nansy++} and other speech-related tasks \cite{huang2023make,huang2022fastdiff,huang2022prodiff,huang2022transpeech,huang2022generspeech}, we consider that fundamental frequency (F0) and timbre are relatively more sensitive to cover versions among different facets since they often change when different artists perform the same piece of song/music. Therefore, we select the F0 and timbre as representatives of the prior knowledge in our work. Specifically, F0 is the musical pitch, representing the high or low notes in the song/music. Timbre describes the vocal characteristics of the artist or instrument, which strongly influences how song/music is heard by trained as well as untrained ears.

\subsection{Framework Overview}
\begin{figure}[htp] \begin{center}
    \includegraphics[width=\linewidth]{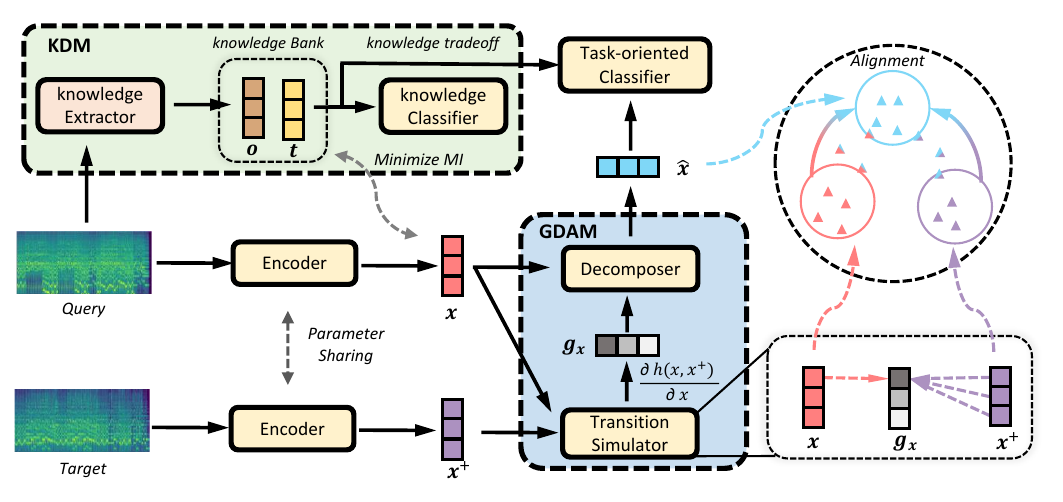}
    \caption{
    	Schematic illustration of DisCover framework. KDM minimizes the MI between the learned representations and version-variant factors that are identified with prior domain knowledge. GADM identifies and decomposes version-variant factors by simulating the representation transitions
        between intra-song versions, and exploits adversarial distillation for effect blocking.
	}
\label{fig:framework}
\vspace{-0.5cm}
\end{center} \end{figure}

To block the intra-version and inter-version biased effects for learning version-invariant representations, we propose DisCover, as shown in Figure \ref{fig:framework}. 
DisCover consists of two modules: (1) Knowledge-guided Disentanglement Module (KDM), which mitigates the negative effect from cover information and extracting the commonness for the versions (green area in the upper left of Figure \ref{fig:framework}). (2) Gradient-based Adversarial Disentanglement Module (GADM), which identifies the differences between versions and alleviates the negative transfer (blue area in the lower right of Figure \ref{fig:framework}). The two modules are jointly trained in a parallel manner.

\subsection{Knowledge-guided Disentanglement} 
As shown in Figure \ref{fig:causal_intervention}(\subref{subfig:causal_intervention_a}), the Knowledge-guided Disentanglement module (KDM) aims to block the bias between intra-song versions (cutoff $Z_i \rightarrow X_i \rightarrow Y_i$), which attempts to make the model more focused on the version-invariant factors $Z$ and learn invariant representations for different cover versions.
Considering that the model is hard to identify the version-specific factors entangled in the representation, as mentioned in Sec \ref{sec:knowledge}, we introduce the prior knowledge (\eg\ F0 and timbre) to serve as the teacher that provides version-variant factors $Z_i$. In contrast to the goal of knowledge transfer, the model aims to minimize the correlation between the learned representations $X_i$ and the version-variant factors $Z_i$.
In this way, we can explicitly disentangle representation $X_i$ from version-variant factors $Z_i$. 

Here, we denote $\boldsymbol{x} \in \mathbb{R}^{dim}$ as the learned representations and $\boldsymbol{z} \in \{\boldsymbol{o}, \boldsymbol{t}\}$ as the knowledge bank of version-variant factors, where $\boldsymbol{o} \in \mathbb{R}^{dim}$ represents the fundamental frequency (F0) features, $\boldsymbol{t} \in \mathbb{R}^{dim}$ represents the timbre representations.
\subsubsection{factors-invariant Representation Modeling}
To minimize the correlation between the learned representations $x$ and the version-variant factors $z$, we introduce mutual information (MI) to serves as the measurement, which is defined as the Kullback-Leibler (KL) divergence between their joint and marginal distributions as: 
\begin{equation}
\begin{aligned}
I(\boldsymbol{x}; \boldsymbol{z}) = \mathbb{E}_{p(\boldsymbol{x}, \boldsymbol{z})}[\log{\frac{p(\boldsymbol{z}|\boldsymbol{x})}{p(\boldsymbol{x})}}]
\end{aligned}
\end{equation}
Since the conditional distribution $p(\boldsymbol{z}|\boldsymbol{x})$ is intractable, we adopt vCLUB \cite{cheng2020club} to approximate the upper bound of MI as:
\begin{equation}
\begin{aligned}
    I(x, z)= \mathbb{E}_{p(x, z)}[\log{q_{\theta_{x,z}}(z|x)}]-\mathbb{E}_{p(x)}\mathbb{E}_{p(z)}[\log{q_{\theta_{x,z}}(z|x)}]
\end{aligned}
\end{equation}
where ${q_{\theta_{x,z}}(\cdot)}$ represents the variational estimation network between $\boldsymbol{x}$ and $\boldsymbol{z}$. Therefore the unbiased estimation for vCLUB between learned representation and version-variant factors can be reformulated as: 
\begin{equation}
\label{eq:l_(z,s)}
\begin{aligned}
    \mathcal{L}_{I(\boldsymbol{x}; \boldsymbol{o})}=\frac{1}{N}\sum_{i=1}^{N}[\log(q_{\theta_{\boldsymbol{x}, \boldsymbol{o}}}(\boldsymbol{o}_i|\boldsymbol{x}_i))-\frac{1}{N}\sum_{j=1}^{N}\log(q_{\theta_{\boldsymbol{x}, \boldsymbol{o}}}(\boldsymbol{o}_j|\boldsymbol{x}_i))]
\end{aligned}
\end{equation}
\begin{equation}
\label{eq:l_(z,t)}
\begin{aligned}
    \mathcal{L}_{I(\boldsymbol{x}; \boldsymbol{t})}=\frac{1}{N}\sum_{i=1}^{N}[\log(q_{\theta_{\boldsymbol{x}, \boldsymbol{t}}}(\boldsymbol{t}_i|\boldsymbol{x}_i))-\frac{1}{N}\sum_{j=1}^{N}\log(q_{\theta_{\boldsymbol{x}, \boldsymbol{t}}}(\boldsymbol{t}_j|\boldsymbol{x}_i))]
\end{aligned}
\end{equation}
where $N$ represents the batch size. By minimizing the Eq. (\ref{eq:l_(z,s)}) and (\ref{eq:l_(z,t)}), we can decrease the correlation between learned representation and version-variant factors and the total MI loss is:
\begin{equation}
\begin{aligned}
    \mathcal{L}_{MI} = \mathcal{L}_{I(\boldsymbol{x}; \boldsymbol{o})} + \mathcal{L}_{I(\boldsymbol{x}; \boldsymbol{t})} 
\end{aligned}
\end{equation}
To obtain the reliable upper bound approximation, a robust variational estimator ${q_{\theta_{x,z}}(\cdot)}$ is required. We train the variational estimator by minimizing the log-likelihood:
\begin{equation}
\begin{aligned}
\mathcal{L}_{q_{\theta_{\boldsymbol{x},\boldsymbol{z}}}}=-\frac{1}{N}\sum_{i=1}^{N}[\log(q_{\theta_{\boldsymbol{x},\boldsymbol{z}}}(\boldsymbol{x}|\boldsymbol{z}))], \boldsymbol{z} \in \{\boldsymbol{o}, \boldsymbol{t}\}
\end{aligned}
\end{equation}

\subsubsection{knowledge tradeoff}
 
However, we argue that vCLUB might be at risk of posterior collapse \cite{he2018lagging, lucas2019don, razavi2018preventing} due to the KL-Vanishinig. For example, if the weights of the variational estimator become randomized due to undesirable training, the introduction of prior knowledge would be meaningless. Therefore, knowledge tradeoff is the self-supervised way to relieve the posterior collapse and ensure training stability for variational estimator.  
Furthermore, considering knowledge extractors' ability, little beneficial version-invariant information might still remain in the  $\boldsymbol{z}$. To address these concerns, we provide two alternatively simple methods. Firstly, we can fuse task-oriented representation $\boldsymbol{e}$ as:

\begin{align}
	a &= \sigma(g(\boldsymbol{e}, q(\boldsymbol{z}))), \\
    \boldsymbol{e}^* &=  a * \boldsymbol{e} + (1 - a) * q(\boldsymbol{z})\label{eq:f0}
\end{align}
where $\sigma(\cdot)$ denotes the sigmoid function, $g(\cdot)$ is the linear transformation, $q(\cdot)$ is the shared MLP in the variational estimator, and $a \in \mathbb{R}$ serves as the 
tradeoff between $\boldsymbol{e}$ and $q(\boldsymbol{z})$. Secondly, we can use clustering models (\eg\ k-means) to annotate the pseudo labels for  $\boldsymbol{z}$ to supervise the  variational estimator with classification task:
\begin{equation}\label{eq:timbre}
\begin{aligned}
\mathcal{L}_{z_{cls}}= -\sum_{i=1}^{N} y_{z_i}log{(\hat{y}_{z_i})}+(1-y_{z_i})log{(1-\hat{y}_{z_i})}
\end{aligned}
\end{equation}
where $y_{z_i}$ is the pseudo label for $z_i$, and $\hat{y}_{z_i}$ is the output of the knowledge classifier.

\subsection{Gradient-based Adversarial Disentanglement}
As shown in Figure \ref{fig:causal_intervention}(\subref{subfig:causal_intervention_b}), the Gradient-based Adversarial Disentanglement module (GADM) aims to block the bias between inter-song versions (cutoff $Z_j \rightarrow X_i \rightarrow Y_i$), which attempts to bridge the intra-group gap and avoid biased representation learning.
As analyzed in \cite{zhou2016learning, selvaraju2017grad, chattopadhay2018grad}, the gradients of the predictive score contain the discriminative information for the downstream tasks. Analogously, the gradients of the transition cost between two versions might convey important information for version-variant factors. For this purpose, we randomly construct the positive query-target pairs with different versions and obtain the corresponding representation pairs $(\boldsymbol{x}, \boldsymbol{x}^+)$ with the same backbone model.    
GADM has three main steps: identification, decomposition, and alignment.

\subsubsection{Identification} 
\label{sec:identification} 
The main idea of identification is to recognize the version-variant factors that are entangled in the elements of learned representations. 
Since the backbone encoder maps the samples into the hyperspace, positive representation pairs $x$ and $x^+$ can be regarded as two points in the same high dimensional space. In the ideal case, different versions of the same song should have similar representations. In other words, these points should cluster together in the hyperspace. However, the distance between two points would be enlarged due to the disruption of version-variant factors that are highly entangled in the representations. Therefore, we treat the distance between query-target pair $x$ and $x^+$ as the transition cost caused by entangled version-variant factors.
Here, we can use metric function (\eg\ Euclidean, Manhattan, or Cosine) to serve as the transitions cost $\mathcal{C}_{trans} \in \mathbb{R}^{+}$ between the representations of intra-song versions $x$ and $x^+$ as:
\begin{equation}
\begin{aligned}
\mathcal{C}_{trans} = h(\boldsymbol{x}, \boldsymbol{x}^+)
\end{aligned}
\end{equation}
where $h(\cdot ,\cdot)$ denotes the metric function. 
Motivated by the GradCAM-like methods \cite{zhou2016learning, selvaraju2017grad, chattopadhay2018grad}, which utilize the saliency-based class information from the gradient perspective. We can obtain version-variant information by calculating the gradients of the transition cost $\mathcal{C}_{trans}$  \textit{w.r.t.} the representation $\boldsymbol{x}$ as: 
\begin{equation}
\begin{aligned}
g_{\boldsymbol{x}}= \frac{\partial\ \mathcal{C}_{trans}}{\partial\ \boldsymbol{x}}
\end{aligned}
\end{equation}
where $g_{\boldsymbol{x}} \in \mathbb{R}^{dim}$ denotes the gradient vector of $\boldsymbol{x}$. 
Since the partial derivative operation for query $\boldsymbol{x}$ utilizes the information from target $\boldsymbol{x}^+$, gradient vector $g_{\boldsymbol{x}}$ probably conveys the element-wise importance information of representation $\boldsymbol{x}$ for measuring the difference to its target $\boldsymbol{x^+}$. Specifically, as shown in the bottom right corner of Figure \ref{fig:framework}, each element $g_{\boldsymbol{x}}(i)$ in $g_{\boldsymbol{x}}$ represents the fusion result between query element $\boldsymbol{x}(i)$ and whole target representation $\boldsymbol{x^+}$. The process allows element $g_{\boldsymbol{x}}(i)$ to automatically search for the elements of the query representation $\boldsymbol{x}$ that are relevant to the transition cost. That's why  $g_{\boldsymbol{x}}$ can identify the version-variant factors hiding in the $\boldsymbol{x}^+$. 
Furthermore, the value of $g_{\boldsymbol{x}}(i)$ represents the sensitivity to the changes of transition cost $\mathcal{C}_{trans}$, where the element with the higher value is more relevant to the version-variant factors based on the nature of gradient.  
\subsubsection{Decomposition} 
After identifying the version-variant factors, we attempt to decompose the version-invariant representation $\hat{\boldsymbol{x}}$ from $\boldsymbol{x}$. Inspired by ToAlign \cite{wei2021toalign}, which decomposes a source feature into a task-relevant/irrelevant one with a gradient-based attention weight vector. We further exploit the numeric order in $g_{\boldsymbol{x}}$ to ensure that the element with the higher gradient has the lower attention weight.
Specifically, given the  gradient vector $g_{\boldsymbol{x}}$, we will construct the corresponding mask vector and decompose it as: 
\begin{equation}
\begin{aligned}
    \boldsymbol{m_x}(i) = \left \{
    \begin{aligned}
        &1 - \frac{\exp{(g_x(i))}}{\sum_{k \in \{k | g_x(k) \geq q_p\}} \exp{(g_x(k))} },\ if\ g_x(i) \geq q_p  \\
        &1,\ otherwise \\
    \end{aligned}
    \right.
\end{aligned}
\end{equation}

\begin{equation}
\begin{aligned}
    \hat{\boldsymbol{x}} = \boldsymbol{m_x} \odot \boldsymbol{x}
\end{aligned}
\end{equation}
where $m_{x}(i)$ denotes $i$-th element in the mask, $q_p$ denotes the $p$-th largest percentile in $g_{\boldsymbol{x}}$, and $\odot$ denotes the hadamard product. 
Moreover, in view of the self-challenging method \cite{huang2020self}, the decomposition process adaptively re-weights $\boldsymbol{x}$ based on the knowledge from $g_{\boldsymbol{x}}$ and forces the backbone to lower the attention on version-specific elements, so as to obtain the version-invariant representation $\hat{\boldsymbol{x}}$.

\subsubsection{Alignment}
To alleviate the negative transfer, we adopt the adversarial distillation sub-module to align entangled representation $\boldsymbol{x}$ to the disentangled one $\hat{\boldsymbol{x}}$. In the beginning,  $\boldsymbol{x}$ and $\hat{\boldsymbol{x}}$ belong to different hyperspheres, where the $\boldsymbol{x}$ is considered as the negative source and the $\hat{\boldsymbol{x}}$ is the positive target. We use them to train the discriminator $D$ to distinguish which hypersphere the representation belongs to, with the classification loss $\mathcal{L}_{D_1}$. Meanwhile, the backbone encoder is trained to fool the discriminator to learn the version-invariant representation by minimizing task-oriented loss while maximizing $\mathcal{L}_{D_2}$:
\begin{equation}
\begin{aligned}
\mathcal{L}_{D_1}=\frac{1}{N}\sum_{i=1}^{N}[\log{D}(\hat{x}_i)+\log{(1-D(x_i)}]
\end{aligned}
\end{equation}

\begin{equation}
\begin{aligned}
\mathcal{L}_{D_2}=\frac{1}{N}\sum_{i=1}^{N}[\log{(1-D(x_i)}]
\end{aligned}
\end{equation}

Furthermore, considering the symmetry of the query-target pair, we can similarly obtain the version-invariant target representation  $\hat{\boldsymbol{x}}^+ \in \mathbb{R}^{dim}$. To ensure the semantic consistency between query and target, it is better to minimize transition cost as:
\begin{equation}
\begin{aligned}
\mathcal{L}_{trans} = h(\hat{\boldsymbol{x}}, \hat{\boldsymbol{x}}^+)
\end{aligned}
\end{equation}

\subsection{Training} 
Given the output of the task-oriented classifier $\hat{y}_{\hat{\boldsymbol{x}}}$, we treat CSI as the classification task, where the task-oriented learning objective can be formulated as follows:
\begin{equation}
\begin{aligned}
\mathcal{L}_{task}= -\sum_{i=1}^{N} y_{\hat{\boldsymbol{x}}}log{(\hat{y}_{\hat{\boldsymbol{x}}})}+(1-y_{\hat{\boldsymbol{x}}})log{(1-\hat{y}_{\hat{\boldsymbol{x}}})}
\end{aligned}
\end{equation}
where $y_{\hat{\boldsymbol{x}}}$ is the groundtruth label. 
To be clear, the overall optimization objective of our proposed DisCover is summarized as follows:
\begin{equation}
\begin{aligned}
\mathcal{L}_1=\mathcal{L}_{task}+\mathcal{L}_{trans}+\lambda_1 \mathcal{L}_{MI}+ \mathcal{L}_{z_{cls}} -\mathcal{L}_{D_2}
\end{aligned}
\end{equation}

\begin{equation}
\begin{aligned}
\mathcal{L}_2=\mathcal{L}_{D_1}+ \lambda_2 \mathcal{L}_{q_{\theta_{\boldsymbol{x},\boldsymbol{z}}}}
\end{aligned}
\end{equation}
where $\mathcal{L}_1$ and $\mathcal{L}_2$ are optimized alternately.

\section{EXPERIMENTS}
We analyze the DisCover framework and demonstrate its effectiveness by answering the following research questions:

\begin{itemize}[leftmargin=*]
	\item \textbf{RQ1}: How does DisCover perform compared with existing best-performing cover song identification methods in different scenarios (\eg, unseen songs/versions) ?
	\item \textbf{RQ2}: Do knowledge-guided disentanglement and gradient-based disentanglement all contribute to the effectiveness over various base models in a model-agnostic manner?
	\item \textbf{RQ3}: How does different architecture and hyper-parameter settings will affect the performance of DisCover?
	\item \textbf{RQ4}: Does DisCover disentangle the version-variant factors?
\end{itemize}

\begin{table}[h]
  \caption{Dataset statics}
  \label{tab:dataset}
  \begin{tabular}{ccccc}
    \toprule
    Dataset & Songs & Recordings & Avg. versions & Language\\
    \midrule
    SHS100K & 10000 & 104641 & 10.5 &English\\
    Karaoke30K & 11500 & 31629 & 2.8 & Chinese\\
    Covers80 & 80 & 160 & 2.0 & English\\
  \bottomrule
\end{tabular} \vspace{-0.5cm}
\end{table}

\subsection{Experimental Setting}

\begin{table*}[h]
\centering
    \caption{Improvement over the best-performing baselines across different scenarios.}
{\setlength{\tabcolsep}{0.6em}
\begin{tabular}{lcccccccccccc}
\toprule
\multicolumn{1}{c}{\multirow{3}{*}{Model}} & \multicolumn{6}{c}{SHS100K} & \multicolumn{6}{c}{Covers80} \\ \cmidrule(lr){2-7} \cmidrule(lr){8-13}
\multicolumn{1}{c}{} & \multicolumn{3}{c}{Scenario 1 :} & \multicolumn{3}{c}{Scenario 2 :} & \multicolumn{3}{c}{Scenario 1 :} & \multicolumn{3}{c}{Scenario 2 :} \\ \cmidrule(lr){2-4} \cmidrule(lr){5-7} \cmidrule(lr){8-10} \cmidrule(lr){11-13}
\multicolumn{1}{c}{}  & MAP↑ & P@10↑ & MR1↓ & MAP↑ & P@10↑ & MR1↓ & MAP↑ & P@10↑ & MR1↓ & MAP↑ & P@10↑ & MR1↓\\ \midrule
2DFM & - & - & - & 0.104 & 0.113 & 415 & - & - & - &0.381 &0.053 &33.60  \\
Ki-CNN &0.176  &0.224  &105.79  &0.215 	&0.183 	&147.3  &0.485 	&0.069 	&16.18  &0.509 &0.071 &15.45  \\
TPPNet & 0.419 & 0.455 & 45.85 & 0.471 & 0.338 & 74.38 & 0.757 & 0.084 & 5.81 & 0.786 & 0.087 & 8.39 \\
CQTNet & 0.571 & 0.573 & \textbf{31.69} & 0.624 & 0.340 & \textbf{61.31} & 0.805 & 0.087 & 6.58 & 0.846 & 0.089 & 5.13 \\
PICKiNet & 0.617 &0.602 &38.66 &0.626 &0.408 &84.12  &0.818 &0.085 &7.11 & 0.858 & 0.091 & 4.27   \\
 \midrule
TPPNet-Dis & 0.565 & 0.567 & 41.60 & 0.561 & 0.384 & 74.26 & 0.814 & 0.091 & 7.81 & 0.849 & 0.091 & 4.74 \\
CQTNet-Dis & \textbf{0.658} & 0.627 & 37.98 & 0.640 & 0.417 & 76.41 & \textbf{0.856} & \textbf{0.091} & \textbf{5.11} & \textbf{0.912} & \textbf{0.095} & \textbf{2.43} \\
PICKiNet-Dis & 0.657 & \textbf{0.627} & 46.80 &\textbf{0.653} &\textbf{0.421} &72.30   &0.830 &0.087 &5.88 & 0.882 & 0.093 & 3.26 \\
\bottomrule
\end{tabular}} \vspace{1ex}\label{tab:comparison}
\end{table*}

\begin{table}[h]
\centering
    \caption{Comparing different methods on Karaoke30K with different scenarios. }
{\setlength{\tabcolsep}{0.4em}\begin{tabular}{lcccccc}
\toprule
\multicolumn{1}{c}{\multirow{2}{*}{Model}} & \multicolumn{3}{c}{Scenario 1 :} & \multicolumn{3}{c}{Scenario 2 :} \\ \cmidrule(lr){2-4} \cmidrule(lr){5-7}
 & MAP↑ & P@10↑ & MR1↓ & MAP↑ & P@10↑ & MR1↓ \\ \midrule
Ki-CNN  &0.483 	&0.119 	&52.53 	&0.524 	&0.116 	&52.01   \\
TPPNet  & 0.760 & 0.165 & 17.65 & 0.777 & 0.154 & 13.86\\
CQTNet  & 0.863 & 0.182 & 7.84 & 0.831 & 0.161 & 11.93\\
PICKiNet  & 0.944 & 0.194 & 4.41 &0.959  &0.178  &4.12\\ \midrule
TPPNet-Dis  & 0.935 & 0.192 & 5.23 & 0.957 & 0.177 & 3.24\\
CQTNet-Dis  & \textbf{0.976} & 0.198 & 2.66 & \textbf{0.983} & \textbf{0.180} & \textbf{3.20} \\
PICKiNet-Dis   & 0.974 & \textbf{0.198} & \textbf{2.61} &0.973  &0.179  &3.52\\ \bottomrule
\end{tabular}} \label{tab:comparison2} \vspace{-0.5cm}
\end{table}

\subsubsection{Dataset}
We conduct experiments on two open source datasets commonly used in cover song identification and one self-collected real-world dataset. Statistics of these datasets are shown in Table \ref{tab:dataset}.
\begin{itemize}[leftmargin=*]
	\item \textbf{Second Hand Songs 100K (SHS100K)}: We downloaded raw audios through youtube-dl\footnote{\url{https://github.com/ytdl-org/youtube-dl}} using the URLs provided on GitHub\footnote{\url{https://github.com/NovaFrost/SHS100K2}}. It has 10000 songs with 104641 recordings. Notablely, there are 25\% of test songs seen during model training in the setting of \cite{yu2019temporal}. To further explore the generalization performance, we also construct another scenario setting where all test songs are unseen during training. For both scenarios, the ratio among the training set, validation set, and testing set is 8:1:1.
	\item \textbf{Covers80}\footnote{\url{https://labrosa.ee.columbia.edu/projects/coversongs/covers80/}}: It has 80 songs with 160 recordings, where each song has 2 cover versions. Due to the small amount of data, it is commonly used only for evaluating models.
	\item \textbf{Karaoke30K}: A real-world Chinese karaoke dataset collected by ourselves. It has 11500 songs with 31629 recordings, where each song has 1 to 3 cover versions. Following the SHS100K, we also construct the two scenarios with the same setting.
\end{itemize}
\subsubsection{Evaluation Metrics}
Following the evaluation protocol of the Mirex Audio Cover Song Identification Contest\footnote{\url{https://www.music-ir.org/mirex/wiki/2020:Audio_Cover_Song_Identification}}, we employ three widely used metrics for evaluation, \ie, MAP (mean average precision), P@10 (precision at 10), and MR1 (mean rank of the first correctly identified cover).
\subsubsection{Comparison Baselines}
\begin{itemize}[leftmargin=*]
    \item \textbf{2DFM} \cite{ellis2012large}: 2DFM transforms a beat-synchronous chroma matrix with a 2D Fourier transformer and poses the search for cover songs as estimating the Euclidean distance.
	\item \textbf{ki-CNN} \cite{xu2018key}: ki-CNN uses a key-invariant
convolutional neural network robust against key transposition
for classification.
	\item \textbf{TPPNet} \cite{yu2019temporal}: TPPNet combines CNN architecture with temporal pyramid pooling to extract information on different scales and transform songs with different lengths into fixed-dimensional representations.
	\item \textbf{CQTNet} \cite{yu2020learning}: CQTNet uses carefully designed kernels and dilated convolutions to extend the receptive field, which can improve the model's representation learning capacity.
	\item \textbf{PICKiNet} \cite{o2021detecting}: PICKiNet devises pitch class blocks to obtain the key-invariant musical features.
\end{itemize}
\subsubsection{Implementation Details}
We train models on the SHS100K and Karaoke30K and report the evaluation metrics on them with different scenarios. Covers80 is used to evaluate the models trained on SHS100K since their languages are the same.  We use parselmouth\footnote{\url{https://github.com/YannickJadoul/Parselmouth}} and resemblyzer\footnote{\url{https://github.com/resemble-ai/Resemblyzer}} to extract F0 and timbre respectively. In KDM, we apply Eq. (\ref{eq:f0}) to F0 feature and Eq. (\ref{eq:timbre}) to timbre representation, where the number of the clusters for generating pseudo label $N =$ 100.  Following the default MI-related setting in \cite{wang2021vqmivc}, we set hyper-parameters $\lambda_1 =$ 0.05, $\lambda_2 =$ 1. In GADM, we select Euclidean distance as the metric function, and the mask ratio is set to 1. Following the setting of \cite{yu2019temporal}, we also apply a multi-length training strategy. Adam \cite{Kingma_Ba_2015} is used as the optimizer for backbone, discriminator, and variational estimator. The training batch size $N$ is 32, initial learning rate is 4e-4, weight decay is 1e-5. Notably, LayerNorm is applied in DisCover to obtain normalized representation to ensure numerical stability in similarity-based retrieval. 

\subsection{Overall Results (RQ1)}
We instantiate the proposed DisCover framework on three best-performing CSI methods, \ie, TPPNet, CQTNet,and PICKiNet, and obtain TPPNet-Dis,  CQTNet-Dis and PICKiNet-Dis. 
Table \ref{tab:comparison} and \ref{tab:comparison2} list the comparison results of the best-performing  models and those enhanced by DisCover on the SHS100K, Karaoke30K and Covers80 datasets under two different scenarios. Specifically, in Scenario \#1, all test songs are unseen during training, while in Scenario \#2, models have seen 25\% class of test songs during training. According to the results, we have the following observations:
\begin{itemize}[leftmargin=*]
    \item Overall, the results across multiple evaluation metrics consistently indicate that TPPNet-Dis, CQTNet-Dis, and PICKiNet-Dis achieve better results than their base models among different datasets and scenarios. Especially, CQTNet-Dis and PICKiNet-Dis show comparable performance and outperform other best-performing methods. We attribute the improvements to the fact that baselines succeed in learning the version-invariant representations by disentangling version-specific musical factors.

    \item  DisCover can boost the performance of models in different scenarios, especially in scenario \#1, where all test songs are unseen. It suggests that the version-variant factors have been highly disentangled. In addition, in Karaoke30k where the cover versions of a particular song are fewer, DisCover could still significantly improve the baselines.
    These results demonstrate the practical merits of DisCover, \ie, identifying version-variant factors with limited number of annotated versions. Note that in real-world scenarios, less popular songs constitute the majority of the music collections and have fewer cover versions.
    In summary, these results demonstrate the strengths of DisCover in generalization and few-shot learning, which is critical for industrial scenarios where music collections could be rapidly updated and too massive to sample the full cover versions for training. 
    
    \item Surprisingly, MR1 scores in scenario \#1 are mostly worse than those in scenario \#2 for all models, especially in SHS100K. These results might suggest that entangled training leads to spurious correlations among songs, including those testing songs seen during training.
    We also observe that on the SHS100K dataset, the proposed method could not beat some baselines \textit{w.r.t.} MR1. SHS100K are known to have unusual audio manifestations in recordings and vocal concert songs  (with strong background noises \eg\ claps, shouts, or whistles), where MR1 scores are sensitive to these noises and exhibit high variances. On the Karaoke30K dataset where the manifestations in recordings are closer to real-world search scenarios, we observe consistent performance improvement brought by DisCover across all metrics.     
   
\end{itemize}

\begin{table}[] \small
\centering
    \caption{Ablation studies by selectively discarding the knowledge-guided disentanglement module (w/o. KDM) and gradient-based adversarial disentanglement module (w/o. GADM). We study both TPPNet-Dis and CQTNet-Dis on different datasets to reveal the model-agnostic capability of the proposed modules.}
{\setlength{\tabcolsep}{0.05em}\begin{tabular}{lcccccccccccc}
\toprule
\multicolumn{1}{c}{Scenario 1 :} & \multicolumn{3}{c}{SHS100K} & \multicolumn{3}{c}{Covers80}  & \multicolumn{3}{c}{Karaoke30K} \\ \midrule
\multicolumn{1}{c}{Model} & MAP↑ & P@10↑ & MR1↓ & MAP↑ & P@10↑ & MR1↓ & MAP↑ & P@10↑ & MR1↓\\ \midrule
TPPNet-Dis & 0.565 & 0.567 & 41.60  & 0.814 & 0.091 & 7.81 & 0.935 	&0.192 	&5.23  \\ 
w/o. KDM & 0.542 & 0.551 & 39.98 &0.805 &0.085 &7.06  & 0.921 & 0.190 & 3.79 \\
w/o. GADM & 0.497 & 0.522 & 48.55 &0.790 &0.087 &8.34  & 0.845 & 0.180 & 8.52 \\ 
TPPNet & 0.419 & 0.455 & 45.85 & 0.757 & 0.084 & 5.81 & 0.760 &0.165  &17.65  \\ \midrule
CQTNet-Dis & 0.658 & 0.627 & 37.98 & 0.856 & 0.091 & 5.11 & 0.976 & 0.198 & 2.66 \\
w/o. KDM & 0.649 & 0.622 & 32.34 &0.843 &0.093 &4.45  & 0.961 & 0.196 & 3.73 \\
w/o. GADM & 0.619 	&0.607 	&36.66 &0.833 &0.088 &7.17 & 0.887 & 0.186 & 7.88 \\ 
CQTNet & 0.571 & 0.573 & 31.69 & 0.805 & 0.087 & 6.58 &0.863 &0.182 &7.84  \\ \bottomrule

\end{tabular}} \label{tab:ablation} 
\end{table}

\begin{table}[h] \small
\centering
\caption{Study of different prior knowledge. The disentanglement of both F0 and timbre can be beneficial.}\label{tab:knowledge}
{\setlength{\tabcolsep}{0.6em}\begin{tabular}{@{}lcccccc@{}}
\toprule
\multicolumn{1}{c}{Scenario 1:} & \multicolumn{3}{c}{SHS100K} & \multicolumn{3}{c}{Covers80} \\ \midrule
\multicolumn{1}{c}{Factors}      & MAP↑      & P@10↑     & MR1↓     & MAP↑     & P@10↑    & MR1↓   \\\midrule
TPPNet                       & 0.420 & 0.454 & 44.33   &0.757 &0.084 &5.81\\ \midrule 
F0                           & 0.457 & 0.485 & 48.20 &0.764 &0.086 	&7.54   \\ 
w/. tradeoff    &0.463 	&0.490 	&43.65  &0.778 	&0.086 	&8.31   \\\midrule %
Timbre                       & 0.443 & 0.475 & 55.75 &0.772 &0.086 	&8.38     \\ 
w/. tradeoff             & 0.466 & 0.495 & 52.69 &0.784 &0.088 &7.14    \\\midrule 
Timbre \& F0                 & 0.469 & 0.496 & 43.96 &0.782 &0.086 &8.73    \\ 
w/. tradeoff & 0.497 & 0.522 & 48.55  &0.790 &0.087 &8.34   \\ 
\bottomrule \toprule 
CQTNet                   & 0.569 & 0.572 & 31.90 &0.805 &0.087 &6.58 \\ \midrule
F0                           & 0.585  & 0.580  & 34.52  &0.814 	&0.093 &4.00      \\ 
w/. tradeoff    &0.603 	&0.597 	&34.46  &0.821 	&0.091 	&3.97 \\\midrule
Timbre                       & 0.586  & 0.585 & 38.17 &0.816 &0.093 &4.59\\
w/. tradeoff &0.606 &0.599 &37.17  &0.829 	&0.094 	&4.53 \\\midrule
Timbre \& F0                 & 0.590 & 0.586 & 37.83  &0.824 &0.089 &5.51   \\
w/. tradeoff & 0.619 & 0.607 & 36.66  &0.833 &0.088 &7.17 \\
\bottomrule
\end{tabular}} 
\end{table}

\subsection{Model Analysis (RQ2, RQ3)}

\subsubsection{Analysis of key building modules.} knowledge-guided disentanglement and gradient-based adversarial disentanglement are two key components of DisCover framework. We conduct the ablation study on them to reveal the efficacy of the architectures and the benefits of disentangling version-variant factors. Specifically, we selectively discard the KDM and GADM from CQTNet-Dis and TPPNet-Dis to obtain ablation architectures, \ie, w/o. KDM, and w/o. GADM, respectively to show the model-agnostic capability of these two modules. The results are shown in Table \ref{tab:ablation}. We can observe that:
\begin{itemize}[leftmargin=*]
	\item Removing either KDM or GADM leads to performance degradation, while removing both modules (\ie, the base model) leads to the worst performance. These results demonstrate the effectiveness of the proposed two modules as well as the benefits of disentanglement for CSI. We attribute this superiority to the fact that the models would absorb less spurious correlations among songs and versions by learning version-invariant representations and blocking intra/inter-version biased effects.
	\item Removing GADM leads to more performance drops than removing KDM, which indicates that introduced prior knowledge only contains the part of the version-variant factors. Therefore it is necessary to identify the remained factors that hide in the representation. These results again verify the effectiveness of the end-to-end disentanglement module GADM.
	\item The results are consistent across different baselines, which indicates that the proposed two modules can easily boost the best-performing CSI baselines in a plug-and-play and model-agnostic manner.
\end{itemize}

\subsubsection{Study of different prior knowledge introduced in KDM} F0 and timbre are two commonly used features in singing voice conversion/synthesis tasks, which can reflect music pitch and voice characteristics, respectively. To further study the impact of different prior knowledge, we selectively use F0 and timbre to serve as the version-variant factors. We conduct experiments on CQTNet and TPPNet with SHS100K dataset. The results are shown in table \ref{tab:knowledge} where we can find that:
\begin{itemize}[leftmargin=*]
    \item Introducing either F0 or timbre can improve the baseline performance and introducing both of them will achieve better results. These results further demonstrate the effectiveness of minimizing the correlation between the learned representations and the version-variant factors.
    \item Different verison-variant factors play different roles in exerting a bad impact on model learning. Compared with F0, disentangling timbre appears to be more beneficial to the baseline models. The reason might be that voice characteristic vary from person to person, which leads to high intra-song variances among versions that are performed by different people.
    \item Learning with knowledge tradeoff leads to better performance with different baselines and datasets, which suggests that this technique can further exploit the useful information hiding in prior knowledge and is helpful in relieving the posterior collapse of variational estimator \cite{he2018lagging, lucas2019don, razavi2018preventing}.  
\end{itemize}

\begin{table}[] \small
\centering
\caption{Analysis of the number of clustering centers N for timbre in knowledge tradeoff on CQTNet and TPPNet under scenario \#1.} \label{tab:clusters}
\setlength{\tabcolsep}{0.6em}\begin{tabular}{cccccccc}
\hline
\multirow{2}{*}{Model} & \multirow{2}{*}{N} & \multicolumn{3}{c}{SHS100K} & \multicolumn{3}{c}{Covers80} \\ \cline{3-8} 
 &  & MAP↑ & P@10↑ & MR1↓ & MAP↑ & P@10↑ & MR1↓ \\ \toprule
\multirow{4}{*}{TPPNet} & 100 & 0.466 & 0.495 & 52.69 &0.784 &0.088 &7.14 \\ 
 & 1K & 0.463 & 0.492 & 43.06 &0.785 &0.088 	&9.83   \\
 & 5K & 0.462 & 0.492 & 44.06  &0.792 &0.084 &8.68 \\
 & 10K & 0.467 & 0.496 & 44.46 &0.777 &0.086 &8.14   \\ \midrule
\multirow{4}{*}{CQTNet} & 100 & 0.606 & 0.599 & 37.17   &0.829 	&0.094 	&4.53\\
 & 1K & 0.601 & 0.594 & 32.39 &0.827 &0.092 &2.95 \\
 & 5K & 0.593 & 0.590 & 38.47 &0.832 &0.096 &3.32  \\
 & 10K & 0.609 & 0.602 & 34.34 &0.833 &0.092 &4.18   \\ \bottomrule 
\end{tabular}
\end{table}

\subsubsection{Analysis of the number of clustering centers for timbre in knowledge tradeoff.}  In this experiment,
we analyze the impact of the number (N) of clusters used to generate pseudo-labels on the model performance, which uncovers the hyper-parameter sensitivity. 
As shown in Table \ref{tab:clusters}, the model performance is overall insensitive to the number of clusters. In other words, the model can achieve comparable performance with relatively few pseudo-labels (\eg\ N $=$ 100) and lower complexity, which is suitable for real-world scenarios to reduce resource consumption.

\begin{table}[] \small
\centering
\caption{Analysis of transition simulation in GADM on CQTNet and TPPNet under scenario \#1.} \label{tab:transition_cost}
\setlength{\tabcolsep}{0.4em}
\begin{tabular}{cccccccc}
\hline
\multirow{2}{*}{Model} & \multicolumn{1}{c}{\multirow{2}{*}{Method}} & \multicolumn{3}{c}{SHS100K} & \multicolumn{3}{c}{Covers80} \\ \cline{3-8} 
 & \multicolumn{1}{c}{} & MAP↑ & P@10↑ & MR1↓ & MAP↑ & P@10↑ & MR1↓ \\ \toprule
\multirow{3}{*}{TPPNet} 
 & Manhattan & 0.344 & 0.389 & 79.87 &0.724 &0.082 &8.58 \\
 & Euclidean & 0.542 &0.551 	&39.98  & 0.805 &0.085 &7.06 \\ 
 & Cosine & 0.473 & 0.495 & 44.08  &0.730 &0.084 &10.35  \\ \midrule 
\multirow{3}{*}{CQTNet} 
 & Manhattan & 0.620 & 0.603 & 34.01 &0.814 &0.089 &4.41 \\ 
 & Euclidean & 0.649 & 0.622 & 32.34 &0.843 &0.093 &4.45 \\ 
 & Cosine & 0.525 & 0.538 & 48.75 &0.784 &0.091 &5.15  \\ \bottomrule 
\end{tabular}
\vspace{-0.5cm}
\end{table}

\subsubsection{Analysis of transition simulation in GADM} As analyzed in Sec. \ref{sec:identification}, we use metric function to serve as the transition cost between two versions of a song. Therefore a reliable metric function is vital for identifying the version-variant factors between different versions. In this experiment, we select three commonly used distances (\eg\ Euclidean, Manhattan, and Cosine) to serve as the transition cost. Surprisingly, as shown in Table \ref{tab:transition_cost}, the Euclidean distance, which is less explored in the CSI literature, shows a clear advantage over other widely used metric functions. This is an interesting finding that might be potentially inspirational. We plan to further uncover the underlying mechanisms in the future.

\begin{figure}[htp] 
\subcaptionbox{CQTNet Baseline\label{subfig:case_study_a}}[0.45\columnwidth]{
\includegraphics[width=0.45\columnwidth]{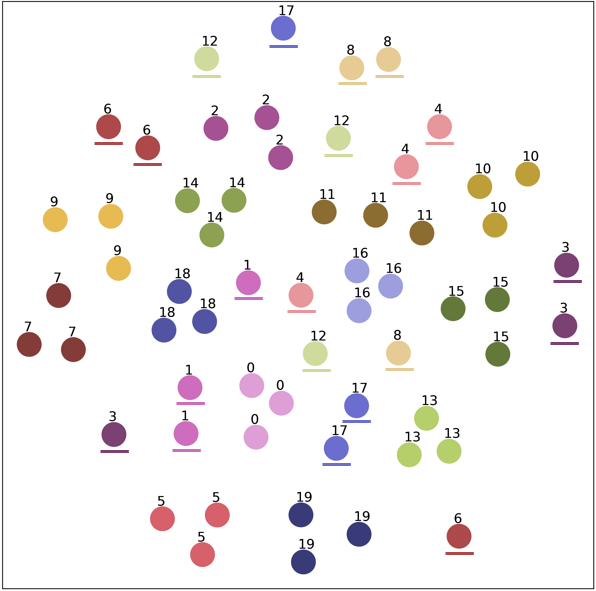}
}\quad
\subcaptionbox{CQTNet with DisCover\label{subfig:case_study_b}}[0.45\columnwidth]{
\includegraphics[width=0.45\columnwidth]{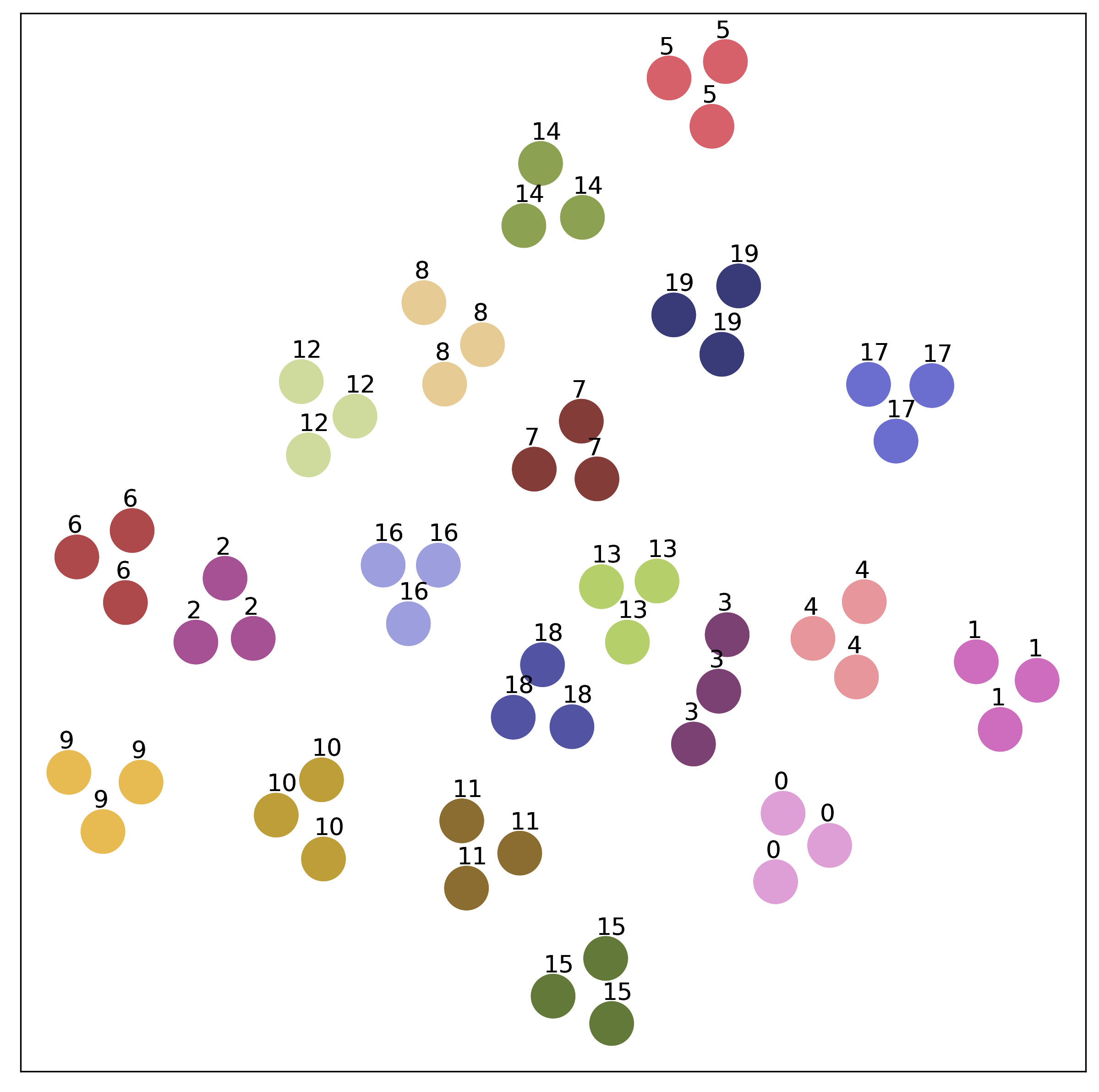}
}
\subcaptionbox{TPPNet Baseline\label{subfig:case_study_c}}[0.45\columnwidth]{
\includegraphics[width=0.45\columnwidth]{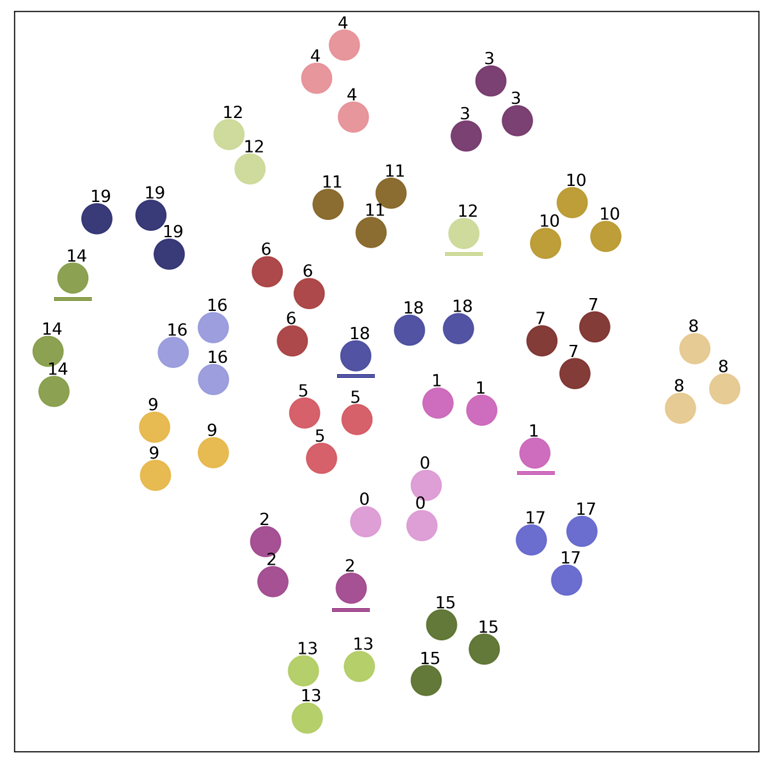}
}\quad
\subcaptionbox{TPPNet with DisCover\label{subfig:case_study_d}}[0.45\columnwidth]{
\includegraphics[width=0.45\columnwidth]{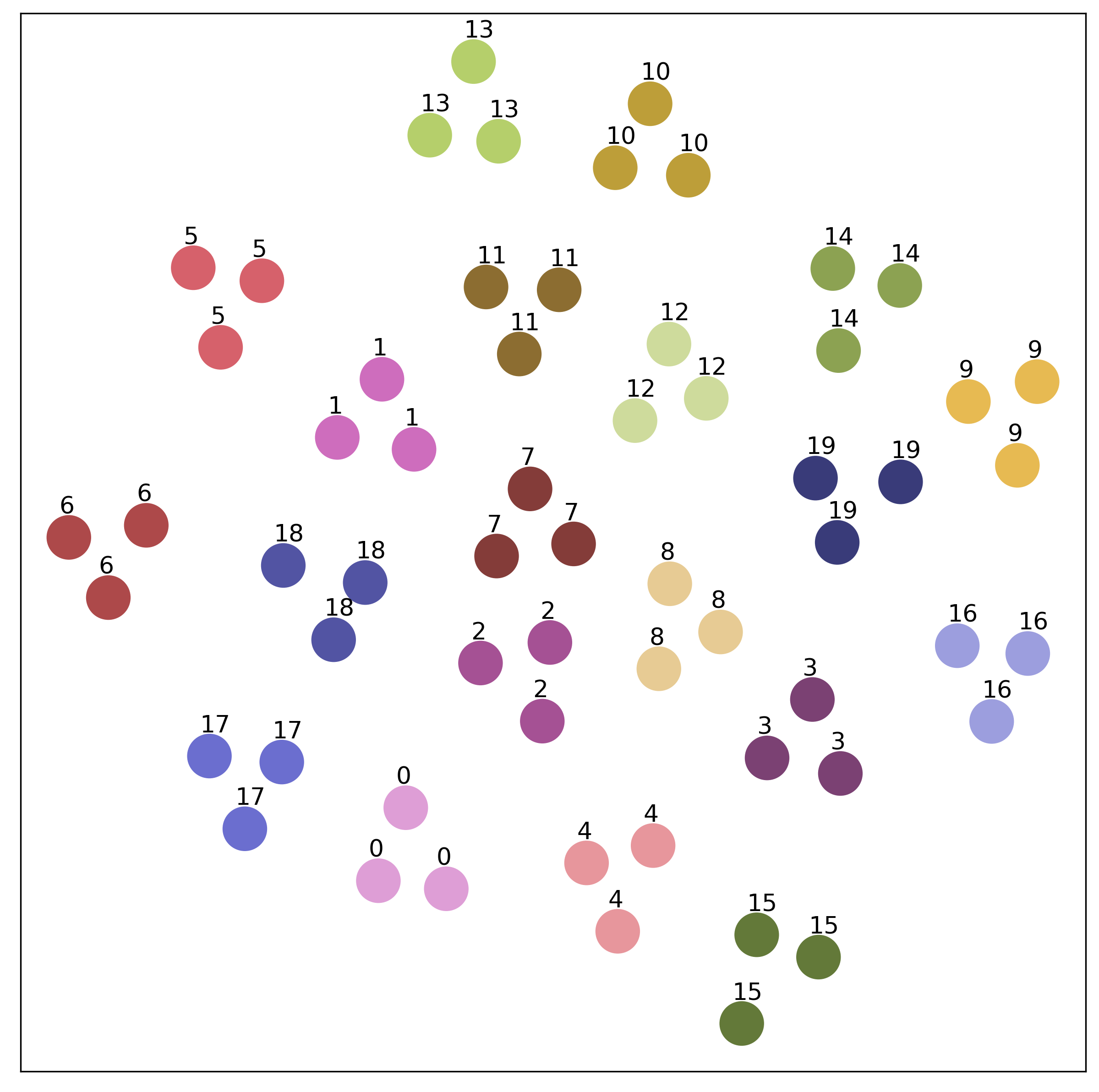}
}
\subcaptionbox{PICKiNet Baseline\label{subfig:case_study_e}}[0.45\columnwidth]{
\includegraphics[width=0.45\columnwidth]{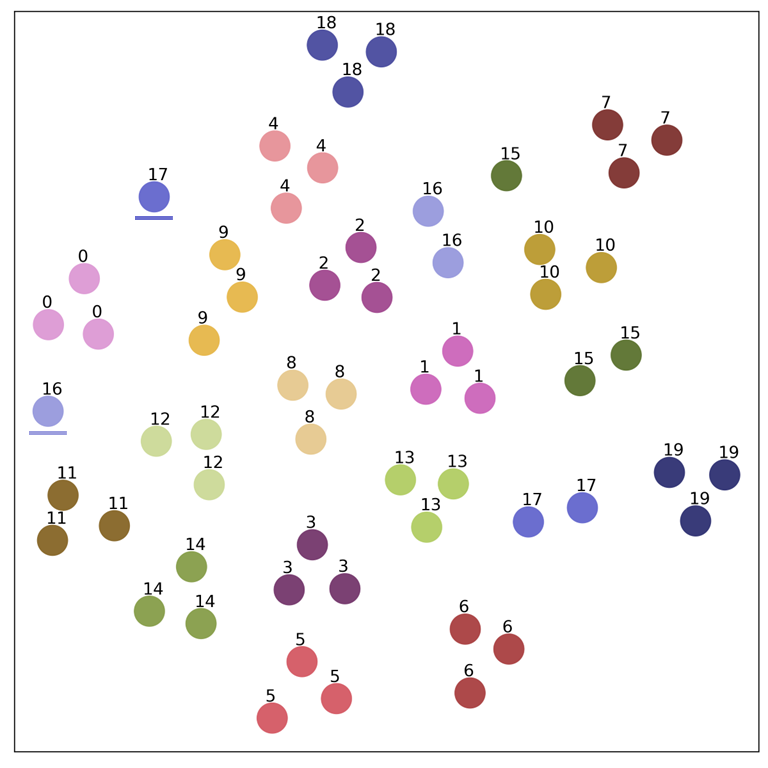}
}\quad
\subcaptionbox{PICKiNet with DisCover\label{subfig:case_study_f}}[0.45\columnwidth]{
\includegraphics[width=0.45\columnwidth]{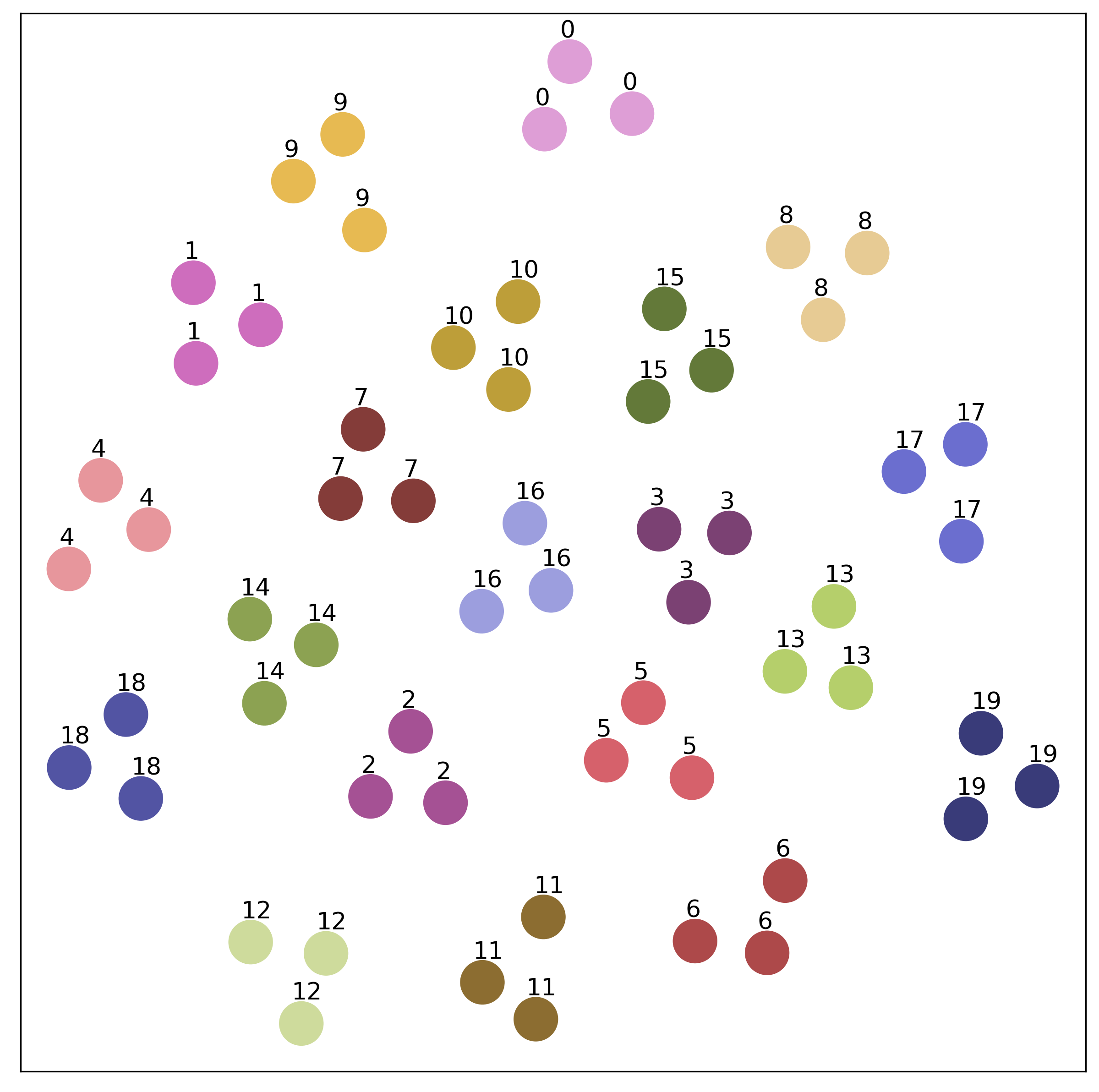}
}
\caption{Case study with t-SNE transformed embeddings derived from different baselines with our DisCover framework, where colored nodes represent the different songs.}
\label{fig:case_study}
\vspace{-0.6cm}
\end{figure} 
\subsection{Qualitative Analysis (RQ4)}
The above analysis quantitatively shows the effectiveness of disentanglement in cover song identification. To evaluate whether the model can learn the version-invariant and unbiased representations via disentangled learning, we visualize the t-SNE transformed embeddings. We adopt CQTNet, TPPNet and PICKiNet as baseline 
and equip them with DisCover framework
and plot the twenty randomly sampled songs and each song has three versions with the representations encoded by the corresponding model. As shown in Figure \ref{fig:case_study}, we can observe that:
\begin{itemize}[leftmargin=*]
    \item Overall,  different versions of a song exhibit more noticeable clusters with the help of DisCover. The base model is more likely to falsely correlate songs based on the similarity of version-variant factors. For example, the versions of song \#12 in Figure \ref{fig:case_study}(\subref{subfig:case_study_a}) are closer to the other songs, which suggests that CQTNet fail to learn the discriminative representation for them. However, in Figure \ref{fig:case_study}(\subref{subfig:case_study_b}), different versions of song \#12 are more compact, which demonstrates the capability of disentanglement.
    \item Moreover, equipped with DisCover, all of  CQTNet, TPPNet and PICKiNet show better performance in learning more discriminative representations compared to the baselines, which further reveals the model-agnostic capability of DisCover.
    \item Furthermore, although the training samples for each song are limited (2 to 3 cover versions for a song), DisCover can still learn the discriminative representations for unseen songs. These results again verify the strengths of DisCover in generalization and few-shot learning.
\end{itemize}

\section{Conclusion}
In this paper, we first analyze the cover song identification problem in a disentanglement view with causal graph. We identify the bad impact of version-variant factors with two effect paths that need to be blocked. Then, we propose the disentangled music representation learning framework DisCover to block these effects. DisCover consists of two modules: (1) Knowledge-guided Disentanglement module, it mitigates the negative effect of cover information and extracts the commonness for the versions, which makes the model more focused on the version-invariant factors and learning invariant representations for different cover versions. (2) Gradient-based Adversarial Disentanglement module, it identifies the differences between versions and alleviates the negative transfer, which bridges the intra-group gap and avoids biased representation learning. Extensive comparisons with best-performing methods and in-depth analysis demonstrate the effectiveness of DisCover and the necessity of disentanglement for CSI.
\section{ACKNOWLEDGEMENTS}
This work was supported in part by the National Key R\&D Program of China under Grant No.61836002, National Natural Science Foundation of China under Grant No. 62222211 and No.62072397. 
We gratefully acknowledge the support of Mindspore\footnote{\url{https://www.mindspore.cn}}, which is a new deep learning computing framework.

\bibliographystyle{ACM-Reference-Format}
\bibliography{citations}










\end{document}